# Anomalous Microfluidic Phonons Induced by the Interplay of Hydrodynamic Screening and Incompressibility

Tsevi Beatus,[1] Roy Bar-Ziv,[1] and Tsvi Tlusty[2]

[1]Department of Materials and Interfaces, The Weizmann Institute of Science, Rehovot, Israel
[2]Department of Physics of Complex Systems, The Weizmann Institute of Science, Rehovot, Israel

We investigate the acoustic normal modes ("phonons") of a 1D microfluidic droplet crystal at the crossover between 2D flow and confined 1D plug flow. The unusual phonon spectra of the crystal, which arise from long-range hydrodynamic interactions, change anomalously under confinement. The boundaries induce weakening and screening of the interactions, but when approaching the 1D limit we measure a marked increase in the crystal sound velocity, a sign of interaction strengthening. This nonmonotonous behavior of the phonon spectra is explained theoretically by the interplay of screening and plug flow.

Microfluidic two-phase flow offers experimental tools to investigate dissipative nonequilibrium dynamics [1–12]. Microfluidic crystals—ordered arrays of water-in-oil droplets driven by flow—are governed by long-range dipolar interactions and exhibit acoustic normal modes, akin to solid-state phonons [11,12]. These interactions share common themes with other systems driven by a symmetry-breaking field, such as dusty-plasma crystals [13,14], vortices in superconductors [15,16], active membranes [17], and nucleoprotein filaments [18,19]. Thus, microfluidic crystals offer a vista, in the linear flow regime, into many-body physics far from equilibrium. Long-range forces, such as the hydrodynamic dipolar force, are known to be radically affected by boundaries and dimensionality. This has been recently shown for two *disordered* systems: Brownian particles confined to 1D and 2D [20–22] and sedimenting particles [23,24].

In this Letter we examine the direct influence of boundaries on the normal modes of an *ordered* many-body system at low Reynolds number (Re $\sim 5 \times 10^{-4}$). We investigated 1D microfluidic droplet crystals under different degrees of confinement ranging from unconfined 2D flow to 1D flow, where the channel is nearly blocked by droplets (plug flow). The interdroplet forces that fall off as $r^{-2}$ in 2D cross over under confinement to decay as $\exp(-2\pi r/W)$ [20,25,26], where the screening length $W$ is the width of the channel. However, close to plug flow, and despite the weakening of interactions due to screening, the magnitude of interdroplet forces increases as $\tan(R/W\pi)$ due to the crystal's incompressibility, $R$ being the droplet radius. This interplay between hydrodynamic screening and incompressibility is reflected in a nonmonotonous behavior of the phonon spectra. Confinement breaks the translational invariance, which is manifested in the breaking of the *x*-*y* antisymmetry of unconfined spectra. Additionally, the approach to incompressibility in the 1D limit implies a divergence of sound velocity and, indeed, we observed its marked increase.

*Experimental setup.*—The microfluidic device (Fig. 1) was fabricated using standard soft lithography and made of poly-dimethyl-siloxane (PDMS) [1,11]. Water droplets formed at a T junction between water and oil channels under continuous flow, emanating with uniform radii $R$ and fixed interdroplet distance $a$. Channel height $h$ was 10 $\mu$m and droplets had a disklike shape, squeezed between the channel floor and ceiling, thereby flowing in 2D. We confined the crystal transversally by narrowing the channel width, varying the confinement parameter, $\gamma \equiv 2R/W$, from 0.1 ($W = 250$ $\mu$m, practically unconfined) to 0.9 ($W = 50$ $\mu$m). Using a manually controlled motorized microscope stage we followed subcrystals of $\sim$40 droplets for $\sim$30 s on average. Power spectra (Fig. 2) were measured by extracting the droplet trajectories and performing Fourier transforms in space and time. The peaks of the power spectrum define the dispersion relation $\omega(k)$ [11].

*Phonon spectra.*—The measured power spectra for different values of the confinement parameter $\gamma$ are shown in Fig. 2. The dispersion relations show acoustic phonons in the crystal. The *unconfined* crystal [Figs. 2(a) and 2(b)] exhibits a sinelike curve that spans the Brillouin zone with a sound velocity $C_s = \partial\omega/\partial k = 165$ $\mu$m s$^{-1}$ [11]. At the edge of the Brillouin zone $k = \pi/a$, waves travel in the

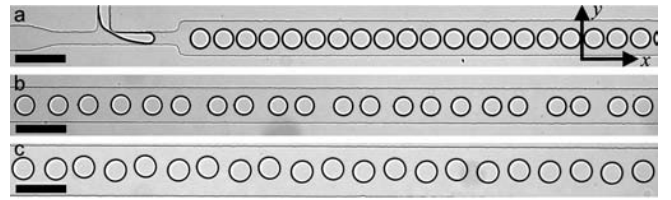

FIG. 1. Droplets of water in oil [mineral oil, viscosity 30 mPa s, with 2% span-80 surfactant (w/w)] were formed at a T junction under constant pressure. Channel height was 10 $\mu$m. (a) Droplet formation in confinement of $\gamma = 0.62$. (b) Longitudinal waves (along *x*) in $\gamma = 0.58$. (c) Transversal waves (along *y*) in $\gamma = 0.46$. Scale bars are 100 $\mu$m.

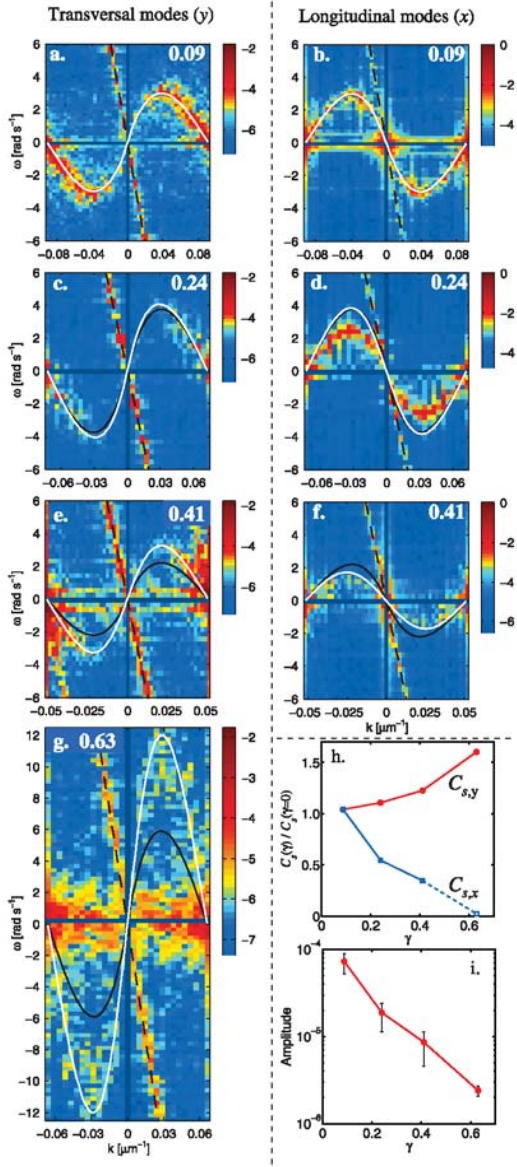

TABLE I. Experimental parameters for Figs. 2(a)–2(g).

| $\gamma$ | $W$ ($\mu$m) | $u_{oil}$ ($\mu$m s$^{-1}$) | $u_d$ ($\mu$m s$^{-1}$) | $C_{s,y}$ ($\mu$m s$^{-1}$) |
|---|---|---|---|---|
| 0.09 | 250 | 1625 | 314 | 165 |
| 0.24 | 150 | 1490 | 308 | 280 |
| 0.41 | 100 | 1310 | 460 | 255 |
| 0.63 | 70  | 1500 | 490 | 700 |

FIG. 2 (color). (a)–(g) Transversal (left column) and longitudinal (right column) phonon power spectra for increasing values of the confinement parameter $\gamma$ (shown on each plot). Color code represents the logarithm of the amplitude. White curves show our theory for $\omega(k)$. Black curves indicate the corresponding theoretical $\omega(k)$ for unconfined crystals. Dashed line is $\omega(k) = -u_d k$. Experimental parameters are shown in Table I. (h) $C_s(\gamma)/C_s(\gamma=0)$ for both polarizations. $C_s(\gamma=0)$ was calculated from theory for an unconfined crystal with the same flow parameters as the confined one. (i) Mean amplitude of the transversal modes as a function of $\gamma$.

opposite direction upon a sign crossover of the group velocity that corresponds to standing waves. The longitudinal and transversal modes are antisymmetric; namely, $\omega_x(k) = -\omega_y(k)$. The straight line $\omega(k) = -u_d k$ results from stationary defects along the channel that deflect the droplets and induce waves that move backwards at $-u_d$ in the crystal frame of reference. Under *confinement*, the general shape of the sinelike curve remained. However, the longitudinal-transversal antisymmetry was broken: $|\omega_x(k)| < |\omega_y(k)|$ for all $k < \pi/a$. This is evident in Fig. 2(h), showing the sound velocities $C_{s,x}$ and $C_{s,y}$, normalized by the theoretically computed unconfined $C_s(\gamma=0)$ [11]. For the longitudinal phonons, $C_{s,x}$ decreased as $\gamma$ increased. Above $\gamma \sim 0.6$ the sinelike curve fell below our detection limit (data not shown). Oppositely, for the transversal phonons $C_{s,y}$ increased with $\gamma$. However, the amplitude of phonons decreased exponentially with $\gamma$ [Fig. 2(i)], essentially undetectable above $\gamma \sim 0.65$. Notably, with the largest $\gamma$ for which we got measurable dispersion [$\gamma = 0.63$, Fig. 2(g)], $C_{s,y}$ (700 $\mu$m s$^{-1}$) was larger than the flow velocity of the crystal itself $u_d$ (490 $\mu$m s$^{-1}$); namely, the flow became "subsonic." This velocity crossover was not observed at smaller $\gamma$. To conclude, under confinement we observed that $C_s$ decreased in the longitudinal modes but increased in the transversal ones. Additionally, the amplitude of vibrations of both modes decreased significantly.

*Hydrodynamic potential.*—We present a hydrodynamic model to account for the effects of confinement. The disklike droplet flows slower than the surrounding oil (due to friction with the channel floor and ceiling), and hence perturbs the flow of oil, which mediates a drag force on the other droplets. The flow far from the droplet can be decomposed into a Poiseuille-like parabolic profile along the $z$ axis (channel height $h$) and a 2D potential flow in the $xy$ plane, such that the potential satisfies the Laplace equation, demanding zero mass flux through the edge of the droplet and the channel sidewalls. The first boundary condition stems from the treadmill flow inside the droplet [3], which implies a nonzero oil velocity parallel to the droplet surface. The solution for an unconfined droplet is a 2D dipole $R^2(u_{oil}^\infty - u_d)\mathbf{r}\hat{x}/r^2$, where $u_{oil}^\infty$ is the oil velocity far from the droplet [11,21,27,28]. As in electrostatics, to achieve zero mass flux through the walls we introduce an array of image dipoles perpendicular to the flow (Fig. 3). Thus, the flow is similar to the flow through a row of pillars [29]. When the droplet is in the middle of the channel at $(x, y) = (0, 0)$ [Fig. 3(a)], the solution is a sum over 2D dipoles positioned on an infinite lattice with a constant $W$. Longitudinal modes do not change the formation of the image lattice that each droplet induces. If, however, the droplet deviates transversally from the middle $(x, y) = (0, \delta)$ [Fig. 3(b)], the array splits into two interlaced arrays of lattice constant $2W$, one displaced by $+\delta$ and the other by $-\delta$. Alternatively, the image array may be considered

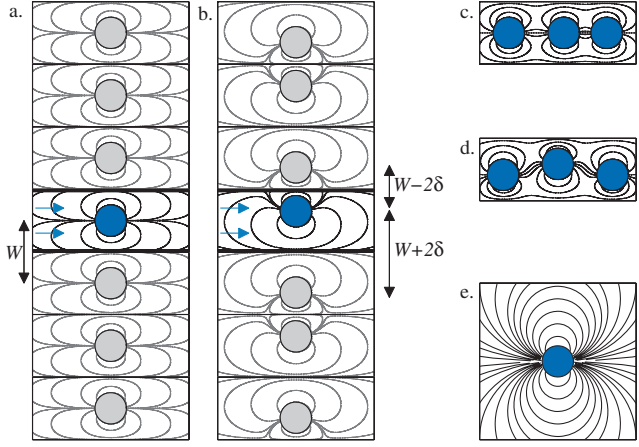

FIG. 3 (color). (a),(b) Flow lines around a single confined droplet (dark blue) are the result of summing the dipole flow fields of its infinite array of reflections (gray). (a) When the droplet is at the center of the channel, the array has uniform spacing $W$. (b) When the droplet is off-centered, the reflections array splits into two interlacing arrays. (c),(d) Flow lines around three droplets with a longitudinal (c) and transversal (d) perturbation. (e) Dipolar flow field of an unconfined droplet.

as a single lattice of constant $2W$ with two droplets per unit cell spaced by $W - 2\delta$. This difference is the source of the $x$-$y$ antisymmetry breaking.

The infinite image array explains why the long-range dipole field is screened under confinement. Far from the droplet, the array of image dipoles can be thought of as a parallel-plate capacitor. Of course, the electrostatic field outside this capacitor vanishes. However, close to the capacitor the discreteness of the charge is significant and the field leaks out and decays exponentially as $\exp(-2\pi r/W)$ [20,30].

To calculate the single-droplet flow potential one sums the contributions of the droplet and its images. It is convenient to use the complex potential $w(z) = \phi_d(z) + i\psi_d(z)$, where $\phi_d$ is the flow potential and $\psi_d$ is the stream function. The total potential is the sum over the interlaced arrays:

$$w_d(z) = R(u_{\text{oil}}^\infty - u_d)\left\{\coth\left[\frac{\pi}{2W}(z - i\delta)\right]\right.$$
$$\left. + \coth\left[\frac{\pi}{2W}(z - i(W - \delta))\right]\right\}$$
$$\times \left[\cot(\pi\gamma/4) - \frac{\sin(\pi\gamma/2)}{\cos(\pi\gamma/2) + \sin(2\pi\delta/W)}\right]^{-1}.$$
(1)

The velocity of oil induced by the droplet is $\nabla\phi_d$, and the drag force that the $j$th droplet exerts on the $i$th droplet is directed along $\nabla\phi_d$ and is given by $\xi_d \nabla\phi_d(\mathbf{r}_i - \mathbf{r}_j)$, with a drag coefficient $\xi_d = 8\pi R^2 \eta / h$ ($\eta$ is oil viscosity). From Eq. (1) we deduce the asymptotic spatial dependence of the force and the effect of confinement:

$$\partial_x\phi \sim (u_{\text{oil}}^\infty - u_d)\begin{cases}\gamma\tan(\pi\gamma/2)\exp(-2\pi x/W) & x \gg W \\ R^2/x^2 & x \ll W.\end{cases}$$
(2)

The behavior in the $y$ direction is similar but the decay length scale is doubled, $W/\pi$, since the image lattice has a double lattice constant (Fig. 3). Thus, there are two competing effects, the exponential spatial decay and the divergence of the amplitude, $\tan(\pi\gamma/2)$, due to confinement. When $\gamma \sim (1 - h/R)$ the exact nature of the divergence may be affected by the boundaries since the flow is no longer a 2D potential flow.

The droplets are subject to a friction force applied by the channel floor and ceiling $F_f = \mu u_d$, with $\mu$ the friction coefficient. The potential of the crystal is approximately a superposition of the single-droplet potentials along with the uniform flow of oil: $\phi(\mathbf{r}) = u_{\text{oil}}^\infty \mathbf{r}\hat{x} + \sum_j \phi(\mathbf{r} - \mathbf{r}_j)$. The equation of motion of the $n$th droplet, considering both drag and friction is, therefore, $\dot{\mathbf{r}}_n = (u_d^\infty/u_{\text{oil}}^\infty)\nabla\phi$, in which we calibrated drag and friction for an isolated droplet moving at $u_d^\infty$ [11].

*Dispersion relations and breaking of antisymmetry.*—The phonon spectra are derived by linearizing the equations of motion for small deviations $(x_n, y_n) \ll a$:

$$\dot{x}_n = -2B\sum_{j=1}^\infty (x_{n+j} - x_{n-j})\coth(\pi j\beta)\text{csch}^2(\pi j\beta),$$
$$\dot{y}_n = \frac{B}{2}\sum_{j=1}^\infty (y_{n+j} - y_{n-j})[3 + \cosh(2\pi j\beta)]\text{csch}^3(\pi j\beta),$$
(3)

where $B \equiv (u_{\text{oil}}^\infty - u_d^\infty)(u_d^\infty/u_{\text{oil}}^\infty)(\pi^2 R/W^2)\tan(\pi\gamma/2)$ and $\beta \equiv a/W$. Substituting propagating plane wave solutions we obtain the dispersion relations:

$$\omega_x(k) = -4B\sum_{j=1}^\infty \sin(jka)\coth(\pi j\beta)\text{csch}^2(\pi j\beta),$$
$$\omega_y(k) = B\sum_{j=1}^\infty \sin(jka)[3 + \cosh(2\pi j\beta)]\text{csch}^3(\pi j\beta).$$
(4)

The computed dispersion relations are superimposed on the experimental data (Fig. 2). The theory recapitulates breaking of the $x$-$y$ antisymmetry $|\omega_x(k)| < |\omega_y(k)|$ for all $k < \pi/a$ and fits the data satisfactorily without any adjustable parameters. The model is somewhat more accurate for transversal modes than for longitudinal ones. Since the amplitude of the modes decreases exponentially with $\gamma$, we could only sample a relatively small part of the predicted behavior of the system, shown in Fig. 4. The figure shows the prediction for the normalized $C_s$ as a function of $\gamma$, calculated for different crystal densities $a/R$. We note that according to this prediction $C_s$ diverges as $\gamma \to 1$ as well as the frequencies of oscillations. In this limit the droplets block the channel (plug flow) and it is difficult to push liquid through the narrow necks, hence the crystal is effectively incompressible. These ''harder'' modes are expected to have reduced amplitudes, which is consistent

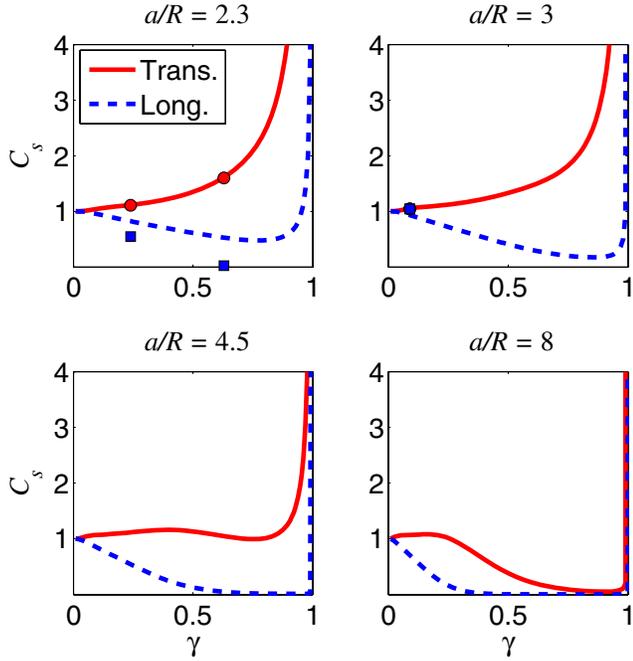

FIG. 4 (color online). Prediction for $C_s$ as a function of $\gamma$ for a few crystal densities $a/R$. The transversal (solid red) curve and the longitudinal $C_s$ (dashed blue) curve are normalized by $C_s(\gamma=0)$ of an unconfined crystal. $C_{s,x}$ and $C_{s,y}$ measured in experiments (Fig. 2) are shown by closed circles and squares, respectively.

with our observations [Fig. 2(i)]. The rich nonmonotonous behavior of $C_s$ stems from the interplay between the exponential decay and the divergence of the amplitude.

The breaking of the $x$-$y$ antisymmetry in confinement originates from the breaking of translation invariance in the $y$ direction. In the unconfined case interdroplet interaction is translation invariant. Therefore, for small displacements $(x_n, y_n)$ $\dot{x}_n \propto \sum_{j \neq n} \partial_x \phi_d(\mathbf{r} - \mathbf{r}_j) \approx \sum_{j \neq n}(x_n - x_j)\partial_x^2 \phi_d((n-j)a)$ and similarly $\dot{y}_n \propto \sum_{j \neq n}(y_n - y_j)\partial_y^2 \phi_d((n-j)a)$. Since $\phi_d$ satisfies the Laplace equation $\partial_x^2 \phi_d = -\partial_y^2 \phi_d$, the equations of motion for $(x_n, y_n)$ are identical up to a sign, thus the dispersion relations are antisymmetric $\omega_x(k) = -\omega_y(k)$. In confinement, however, where $\phi_d$ depends also on the distance from the sidewalls, the equations of motion of $x$ and $y$ are different [Eqs. (3) and (4)]; hence, $\omega_x(k) \neq -\omega_y(k)$.

To summarize, confinement of the 1D droplet crystal enables one to observe a crossover between long-range and screened hydrodynamic forces. The approach to incompressibility and the divergence of sound velocity due to blocking of the channel competes with the screening effect to yield the observed rich phonon spectra. Investigation of microfluidic phonons in other geometries may lead to other interesting and unexplored spectra. The effect of boundaries on long-range forces has also practical implications for the design of droplet-generating devices.

We wish to thank E. Moses, H. Diamant, D. Sapir, S. Safran, and V. Steinberg. This work was supported by a grant from the Israel Science Foundation.